\documentstyle[epsfig,longtable]{aipproc}
\def\ltsima{$\; \buildrel < \over \sim \;$}
\def\simlt{\lower.5ex\hbox{\ltsima}}

\begin{document}
\title{Accretion-Ejection Instability and a ``Magnetic Flood'' 
scenario\\
for GRS 1915+105}

\author{Michel Tagger$^*$}
\address{$^*$Service d'Astrophysique (CNRS URA 2052)\\
C.E.A. Saclay, 91191 Gif sur Yvette (France)
%
\footnote{
e-mail tagger@cea.fr
\begin{flushright}
    \emph{Proceedings of the 5th Compton Symposium, 
    Portsmouth, Sept. 1999}
\end{flushright}}
}

\maketitle

\begin{abstract}
We present an instability, occurring in the inner region of magnetized
accretion disks, which seems to be a good candidate to explain the
low-frequency QPO observed in many X-ray binaries.  We then briefly show
how, in the remarkable case of the microquasar GRS 1915+105, identifying
this QPO with our instability leads to a scenario for the $\sim$ 30 mn
cycles of this source. In this scenario the cycles are controlled by 
the build-up of magnetic flux in the disk.
\end{abstract}

\section*{Introduction}

This contribution comes from two different lines of work: the first one 
is purely theoretical, and has led us to present recently \cite{tagger:TP99} 
an instability which may occur in the inner region of magnetized disks. 
We have called it Accretion-Ejection Instability (AEI), because one of its 
main characteristics is to extract energy and angular momentum from the 
disk, and to emit them {\it vertically} as Alfven waves propagating 
along magnetic field lines threading the disk. These Alfven waves then 
may deposit the energy and angular momentum in the corona above the 
disk, providing an efficient way to energize winds or jets from the 
accretion energy.\\

The second approach has consisted in the comparison of this instability
with the observed properties of the low-frequency (.5 - 10 Hz)
Quasi-Periodic Oscillation (QPO) observed in the low and hard state of
the micro-quasar GRS 1915+105.  The very large and fast growing number of
observational results on this source gives access to many aspects of the
physics of the disk.  They allow this comparison to rely on basic
properties of the instability, and on more detailed ones, such as the
correlation between the QPO and the evolution of the disk and coronal
emissions (identified respectively as multicolor black body and
comptonized power-law tail in the X-ray spectrum).\\

This comparison encourages us to consider that the AEI may indeed be the
source of the QPO. Thus we proceed by considering the $\sim$ 30 mn
cycles of GRS 1915+105.  These cycles are the most spectacular in the
gallery of behaviors and spectral states of this source, in particular
because multi-wavelength observations have shown IR and radio outbursts
coinciding with them, consistent with the synchrotron emission from an
expanding cloud ejected at relativistic speeds.  These
cycles have been analyzed in great details, and the QPO shows a very
characteristic and reproducible behavior.  We have thus built a
scenario, starting from the identification of the QPO with the AEI, and
considering how this could explain the evolution of the source during
the cycle.  We refer to it as a {\it magnetic flood} scenario, because
we are led to believe that the cycle is controlled by the build-up of
the vertical magnetic flux stored within the disk.  The scenario is
compatible with the available information on this type of cycle,
explains a number of results in existing data, and leads to intriguing
considerations on the behavior of GRS 1915+105.

\section*{Accretion-Ejection Instability}

We will present the instability here only in general terms, and refer to
our recent publication \cite{tagger:TP99} for detailed derivation and
results.  It appears in disks threaded by a vertical magnetic field, of
the order of equipartition with the gas pressure ($\beta=8\pi
p/B^2\simlt 1$).  The instability appears essentially as a spiral
density wave in the disk, very similar to the galactic ones, but driven
by the long-range action of magnetic stresses rather than self-gravity. 
The main difference lies in the amplification mechanism: instability
results from the coupling of the spiral density wave with a {\it Rossby
wave}.  Rossby waves, associated with a gradient of vorticity, are 
best known in planetary atmospheres, including the other GRS -- the
Great Red Spot of Jupiter.  In the present case differential rotation
allows the spiral wave to grow by extracting energy and angular momentum
from the disk, and transferring them to a Rossby vortex at its
corotation radius.  This radius is constrained by the physics of the
spiral to be a few times ($\sim 2-5$ for the azimuthal wavenumber $m=1$,
{\it i.e.} a one-armed spiral) the inner radius of the disk.\\

A third type of wave completes the description of the instability: it is
an Alfven wave, emitted along the magnetic field lines towards a
low-density corona above the disk.  The mechanism here is simply that
the Rossby vortex twists the footpoints of the field lines in the disk. 
This twist will then propagate upward, carrying to the corona the energy
and angular momentum extracted from the disk by the spiral.\\

\begin{figure}[] 
\centerline{\epsfig{file=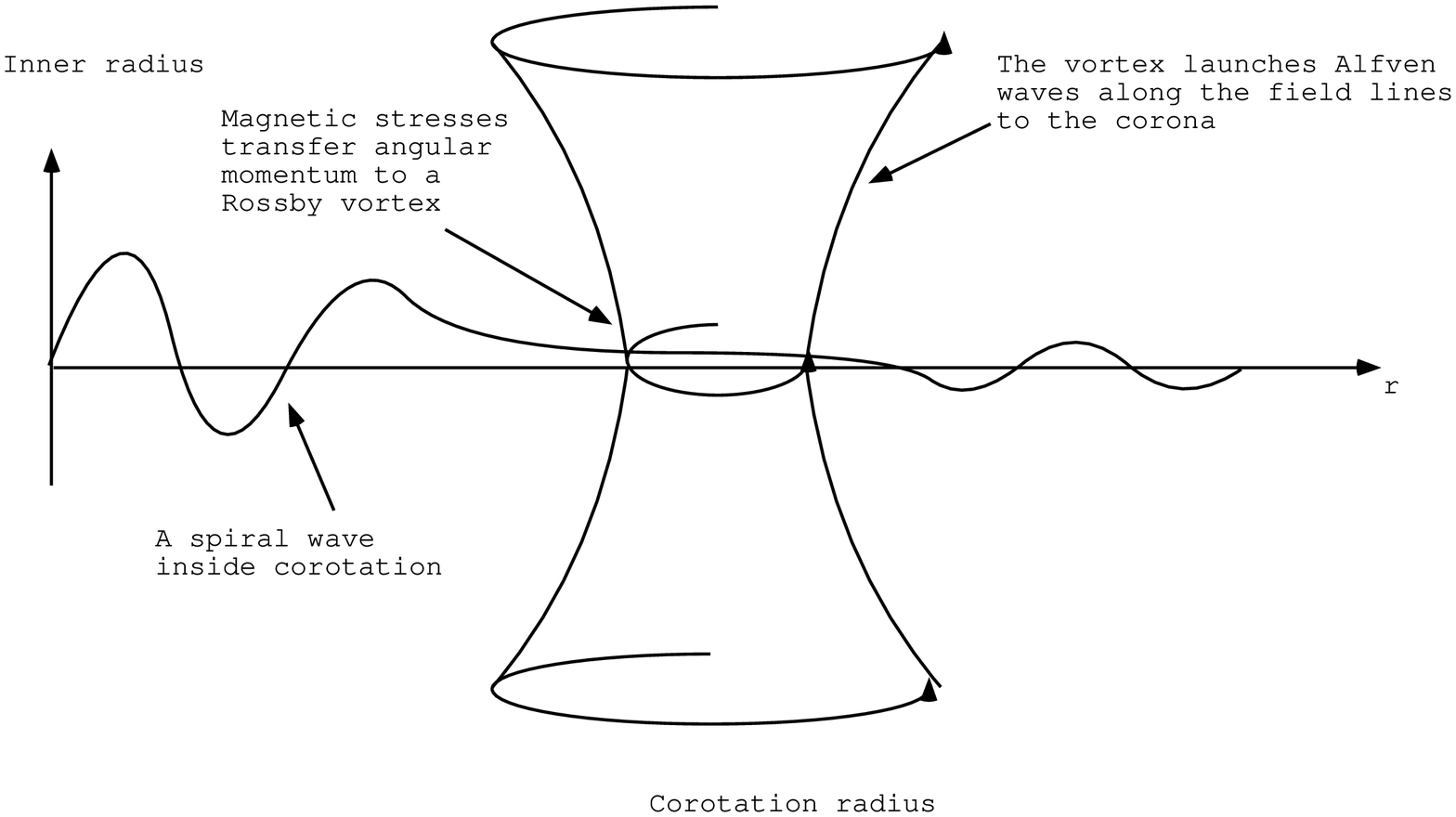, width=\columnwidth,height=2.5in}}
\caption{The propagation of the wave in the disk and its corona. A 
spiral wave grows by extracting energy and angular momentum from the 
disk, and transferring them to a Rossby vortex. The latter in turn 
transfers them to an Alfven wave toward the corona.}
\label{tagger:fig1}
\end{figure}

The mechanism is thus quite complex; this comes essentially from
differential rotation, which allows a mixing of waves which would
otherwise evolve independently.  It results in an instability, growing
on a time scale of the order of $r/h$ times the rotation time (where $h$
is the disk thickness and $r$ its radius).  We will present here its
main characteristics, which will be essential in what follows:

    $\bullet$ It occurs when the vertical magnetic field $B_{0}$ is near
    equipartition ($\beta\simlt 1$) and presents a moderate or strong
    radial gradient. 
    
    $\bullet$ The efficiency of the coupling to the Rossby
    wave selects modes with low azimuthal wavenumbers (the number of
    arms of the spiral), $m=2$ or \hbox{$m=1$} usually, depending on a
    number of parameters (density and temperature profiles, field
    strength, etc.)  
    
    $\bullet$ For a given $m$, the mode frequency is close
    to $\omega\approx (m-1)\Omega_{int}$, the rotation frequency at the
    inner disk radius.  In the special case of the $m=1$ mode, the
    frequency is usually of the order of $\sim .2 - .5$ times
    $\Omega_{int}$.  
    
    $\bullet$ By analogy with galactic spirals, we can
    expect that these properties result in the formation of a large
    scale, quasi-stationary spiral structure rather than in a turbulent
    cascade to small scales.  
    
    $\bullet$ This should strongly affect the
    structure of the disk.  Indeed, underlying the usual model of
    turbulent viscous transport in a disk (leading to Shakura and
    Sunyaev's model of $\alpha$ disks) is the assumption of small scale
    turbulence.  This leads to a {\it local} deposition of the accretion
    energy, efficiently heating the disk.  Here on the other hand, the
    accretion energy is transported away by {\it waves}: extracted from
    the disk by the spiral wave, it is first transferred to the Rossby
    vortex, then to Alfven waves.  Thus, here as in galaxies, the
    connection between gas accretion and disk heating is not as 
    straightforward as in $\alpha$-disks.

\section*{Magnetic flood scenario}

The low-frequency QPO in GRS 1915+105 has been the object of many recent
studies \cite{tagger:swank,tagger:muno}.  During the $\sim 30$ mn cycles
of this source, the QPO appears only during the low state, and its
frequency varies in a repetitive manner during that phase.  Let us
convert its frequency $\nu_{QPO}$ to a keplerian radius $r_{QPO}$, and
compare it to the color radius $r_{color}$ resulting from a multi-color
black body model of the disk emission: observations show that the ratio
$r_{QPO}/r_{color}$ remains of the order of 5 while both radii vary
during the low state.  It is usually considered that $r_{color}$ gives a
measure of the internal radius $r_{int}$ of the disk, although the ratio
$r_{color}/r_{int}$ is subject to some uncertainties.  It is thus very
tempting to consider that the QPO originates from a pattern in the disk,
rotating at a frequency corresponding to a radius $r_{QPO}$ of the order
of a few times $r_{int}$.  This may be supported by a correlation, found
between $\nu_{QPO}$ and a higher frequency feature in various binary
systems, including neutron star and black hole
binaries\cite{tagger:psaltis}.  Although the evidence is fragile in the
case of GRS 1915+105, it would lead to consider that the ratio
$r_{QPO}/r_{int}$ is of the order of 5, in agreement with the previous
result, and corresponding to the value we expect for the $m=1$ AEI.
This, and more detailed arguments to be presented elsewhere, leads us to
tentatively identify the AEI as the source of the QPO, and to consider
how this could fit with the 30 mn cycles of this source.

We start from the conditions responsible for the onset of the
instability, {\it i.e.} a change in $B_{0}$ or its radial gradient.  We
find better agreement with the former, and in this case the sudden
transition from the ``high and soft'' state to the ``low and hard'' one
would find a natural explanation: one has to remember that the best
candidate to explain accretion in a magnetized disk is the
magneto-rotational instability (MRI) \cite{tagger:balbus}.  It appears in disks
with low magnetization ($\beta> 1$), and results in small-scale
turbulence which causes viscous accretion, in agreement with a standard
$\alpha$ disk.

Let us consider that in the high state the disk extends down to its last
stable orbit at $r_{LSO}$, as suggested by the consistent minimal value
of $r_{color}$, and that accretion is caused by the MRI, following an
$\alpha$ prescription.  The MRI might be responsible for the
``band-limited noise'' observed in power density spectra (below the QPO
frequency, {\it i.e.} farther in the disk, when the QPO is present). 
Although numerical simulations of the MRI give estimates of the
resulting $\alpha$, {\it i.e.} turbulent viscosity, they are not able at
this stage to give the associated turbulent magnetic diffusivity, so
that the evolution of the magnetic flux in the disk cannot be
prescribed.  Our main assumption is that in these conditions vertical
magnetic flux builds up in the disk: either because it is dragged in
with gas flowing from the companion, or from a dynamo effect
\cite{tagger:brandenburg}.  This is actually the configuration observed near
the center of the Galaxy.

Then the field must grow in the disk, so that $\beta$ decreases until it
reaches $\beta\simeq 1$, at which point the MRI stops and our
instability sets in, appearing as the low-frequency QPO. The most
important consequence is that turbulent disk heating stops, so that the
disk temperature should drop, further reducing $\beta$.  The abrupt
transition from the high to the low state thus finds a natural
explanation, as a sharp transition between a low magnetization,
turbulently heated state to a high
magnetization one, where disk heating stops and accretion energy is
redirected toward the corona 
(although estimating what fraction of this energy is actually deposited
in the corona depends on the physics of Alfven wave damping).

The content of the space between the disk (when it does not extend down
to $r_{LSO}$) and the black hole is not known.  It might be an ADAF, or
a large-scale, force-free magnetic configuration holding the magnetic
flux frozen in the black hole (following the Blandford-Znajek
mechanism).  In both cases the condition which determines the inner disk
radius $r_{int}$ must be complex, but it is reasonable to assume that a
drop in the disk pressure could explain the increase of $r_{color}$ at
the onset of the low state.  Continuing accretion from the outer disk
region would then move the disk back toward the last stable orbit, as
seen during the low state.

The light curves show an ``intermediate spike'', halfway through the low
state.  At this time $r_{color}$ is back to its minimal value, the QPO
stops, and the coronal emission decreases sharply.  This is also when
the infra-red synchrotron emission, presumably from a ``blob'' ejected
at relativistic speed, begins \cite{tagger:mirabel}.  It is then natural
to consider that at this time a large-scale magnetic event, possibly
reconnection with the magnetic flux surrounding the black hole, causes
ejection of the coronal plasma.  This allows the disk to return to a
lower magnetization state, so that once it has fully recovered it can
start a new cycle in the high and soft state.

\section*{Conclusion}

The properties of the low-frequency QPO in GRS 1915+105 have led us to
tentatively identify it with the Accretion-Ejection Instability.  This
has allowed us to build up a scenario for the 30 mn cycles of this
source.  In contrast with global descriptions, such as $\alpha$ disks,
this does not allow us to predict specific spectral signatures: in the
same manner, knowledge of Rossby waves would hardly allow one to predict
the existence and appearance of the Great Red Spot on Jupiter.  On the
other hand, the scenario is qualitatively compatible with all the
information we have about these cycles.  It explains why and how the QPO
appears, how its frequency varies with the color radius, and why the
transition from the high to the low state has to be a sharp one.  Future
work will be devoted to the QPO behavior at other times in GRS 1915+105,
and then to other sources (black hole or neutron star binaries) where
the identification of the QPO might give access to additional physics.

\end{document}